\RequirePackage{fix-cm}
\documentclass[smallextended]{svjour3}       
\smartqed  
\usepackage{graphicx}
\usepackage{xcolor}
\usepackage[hyphens]{url}
\usepackage[shortlabels]{enumitem}

\usepackage{natbib}
\setcitestyle{authoryear}

\usepackage{IEEEtrantools}
\usepackage{amssymb,amsmath}
\newcommand{\mr}{\mathrm}

\usepackage{tabularx}

\newcommand{\languages}{Arabic, (Mandarin) Chinese, English, French, Italian, Japanese, Korean, Persian, Portuguese, Spanish, Russian and Turkish}

\newcommand{\totalResponses}{2668 }
\newcommand{\invalidResponses}{439 }
\newcommand{\blankResponses}{4 }
\newcommand{\validResponses}{2225 }
\newcommand{\numberOfCountries}{53 }
\journalname{Empirical Software Engineering}
\begin{document}

\title{Pandemic Programming
}
\subtitle{How COVID-19 affects software developers and how their organizations can help}

\author{Paul Ralph          \and
        Sebastian Baltes    \and
        Gianisa Adisaputri  \and
        Richard Torkar      \and
        Vladimir Kovalenko  \and
        Marcos Kalinowski   \and
        Nicole Novielli     \and
        Shin Yoo            \and
        Xavier Devroey      \and
        Xin Tan             \and
        Minghui Zhou        \and
        Burak Turhan        \and
        Rashina Hoda        \and
        Hideaki Hata        \and
        Gregorio Robles     \and
        Amin Milani Fard    \and
        Rana Alkadhi        
}

\institute{
    P. Ralph, Dalhousie University, \email{paulralph@dal.ca} \\            
    S. Baltes, The University of Adelaide, \email{sebastian.baltes@adelaide.edu.au} \\
    G. Adisaputri, Adisa Emergency and Disaster Management, \email{gianisaa@gmail.com} \\
    R. Torkar, Chalmers and University of Gothenburg, and Stellenbosch Institute for Advanced Study \email{richard.torkar@cse.gu.se} \\
    V. Kovalenko, JetBrains \email{vladimir.kovalenko@jetbrains.com} \\
    M. Kalinowski, Pontifical Catholic Uni. of Rio de Janeiro, \email{kalinowski@inf.puc-rio.br} \\
    N. Novielli, University of Bari Aldo Moro \email{nicole.novielli@uniba.it} \\
    S. Yoo, KAIST, \email{shin.yoo@kaist.ac.kr} \\
    X. Devroey, Delft University of Technology \email{X.D.M.Devroey@tudelft.nl} \\
    X. Tan, Peking University, \email{tanxin16@pku.edu.cn} \\
    M. Zhou, Peking University, \email{zhmh@pku.edu.cn} \\
    B. Turhan, Monash University \& University of Oulu, \email{burak.turhan@monash.edu} \\
    R. Hoda, Monash University \email{rashina.hoda@monash.edu} \\
    H. Hata, Nara Institute of Science and Technology \email{hata@is.naist.jp} \\
    G. Robles, Universidad Rey Juan Carlos \email{grex@gsyc.urjc.es} \\
    A. Milani Fard, New York Institute of Technology \email{amilanif@nyit.edu} \\
    R. Alkadhi, King Saud University \email{ralkadi@ksu.edu.sa} \\
}

\date{Received: date / Accepted: date}

\maketitle

\begin{abstract} 
\emph{\newline Context.} As a novel coronavirus swept the world in early 2020, thousands of software developers began working from home. Many did so on short notice, under difficult and stressful conditions.
\emph{Objective.} This study investigates the effects of the pandemic on developers' wellbeing and productivity.
\emph{Method.} A questionnaire survey was created mainly from existing, validated scales and translated into 12 languages. The data was analyzed using non-parametric inferential statistics and structural equation modeling.
\emph{Results.} The questionnaire received \validResponses usable responses from \numberOfCountries countries. Factor analysis supported the validity of the scales and the structural model achieved a good fit ($\mr{CFI} = 0.961$, $\mr{RMSEA} = 0.051$, $\mr{SRMR} = 0.067$). Confirmatory results include: (1) the pandemic has had a negative effect on developers' wellbeing and productivity; (2) productivity and wellbeing are closely related; (3) disaster preparedness, fear related to the pandemic and home office ergonomics all affect wellbeing or productivity. Exploratory analysis suggests that: (1) women, parents and people with disabilities may be disproportionately affected; (2) different people need different kinds of support.   
\emph{Conclusions.} To improve employee productivity, software companies should focus on maximizing employee wellbeing and improving the ergonomics of employees' home offices. Women, parents and disabled persons may require extra support.

\end{abstract}

\section{Introduction}
\label{intro}
In December 2019, a novel coronavirus disease (COVID-19) emerged in Wuhan, China. While symptoms vary, COVID-19 often produces fever, cough, shortness of breath, and in some cases, pneumonia and death. By April 30, 2020, The World Health Organization (WHO) recorded more than 3 million confirmed cases and 217,769 deaths \citep{world2020coronavirus}. With wide-spread transmissions in 214 countries, territories or areas, the WHO declared it a \textit{Public Health Emergency of International Concern}~\citep{who2020statement} and many jurisdictions declared states of emergency or lockdowns~\citep{Kaplan2020lockdown}. Many technology companies told their employees to work from home \citep{Duffy2020Big}.

Thinking of this situation as a global natural experiment in working from home---the event that would irrefutably verify the benefits of working from home---would be na\"ive. This is not normal working from home. This attempting to work from home, unexpectedly, during an unprecedented crisis. The normal benefits of working from home may not apply \citep{donnelly2015disrupted}. Rather than working in a remote office or well-appointed home office, some people are working in bedrooms, at kitchen tables and on sofas while partners, children, siblings, parents, roommates, and pets distract them. Others are isolated in a studio or one-bedroom apartment. With schools and childcare closed, many parents juggle work with not only childcare but also home schooling or monitoring remote schooling activities and keeping children engaged. Some professionals have the virus or are caring for ill family members.

\bigskip
\noindent\begin{tabular}{|p{0.95\textwidth}|}
\hline
\\
\textit{Quarantine work !== Remote work. I've been working remotely with success for 13 years, and I've never been close to burn out. I've been working quarantined for over a month and I'm feeling a tinge if burn out for the first time in my life. Take care of yourself folks. Really.}\\
\hfill\vadjust{}--Scott Hanselman (@shanselman), April 20, \citeyear{Hanselman2020tweet}\\
\\
\hline
\end{tabular}
\bigskip

While numerous studies have investigated remote work, few investigate working from home during disasters. There are no modern studies of working from home during a pandemic of this magnitude because there has not been a pandemic of this magnitude since before there was a world wide web. Therefore, software companies have limited evidence on how to support their workers through this crisis, which raises the following research question.

\smallskip
{\noindent \textit{\textbf{Research question:} How is working from home during the COVID-19 pandemic affecting software developers' emotional wellbeing and productivity?}\par}
\smallskip

To address this question, we generate and evaluate a theoretical model for explaining and predicting changes in wellbeing and productivity while working from home during a crisis. Moreover, we provide recommendations for professionals and organizations to support employees who are working from home due to COVID-19 or future disasters. 


\section{Background}
\label{sec:background}

To fully understand this study, we need to review several areas of related work: pandemics, bioevents and disasters; working from home; productivity and wellbeing. 

\subsection{Pandemics, bioevents and disasters}
\label{sec:pandemics}

\citet{madhav2017chapter} defines pandemics as ``large-scale outbreaks of infectious disease over a wide geographic area that can greatly increase morbidity and mortality and cause significant economic, social, and political disruption'' (p.~35). Pandemics can be very stressful not only for those who become infected but also for those caring for the infected and worrying about the health of themselves, their families and their friends \citep{kim2015public,prati2011social}. In a recent poll, ``half of Canadians (50\%) report[ed] a worsening of their mental health'' during the COVID-19 lockdown \citep{ARI2020worry}. In Australia, the pandemic appears to have doubled the incidence of mental health problems \citep{fisher2020mental}. 

A pandemic can be mitigated in several ways including \textit{social distancing} \citep{anderson2020will}: ``a set of practices that aim to reduce disease transmission through physical separation of individuals in community settings'' \citep[p. 120-14]{rebmann2009infectious}, including public facility shutdowns, home quarantine, cancelling large public gatherings, working from home, reducing the number of workers in the same place at the same time and maintaining a distance of at least 1.5--2m between people \citep{rebmann2009infectious,anderson2020will}. 

The extent to which individuals comply with recommendations varies significantly and is affected by many factors. People are more likely to comply when they have more self-efficacy; that is, confidence that they can stay at home or keep working during the pandemic, and when they perceive the risks as high \citep{teasdale2012importance}. However, this ``threat appraisal'' depends on: the psychological process of quantifying risk, sociocultural perspectives (e.g. one's worldview and beliefs; how worried one's friends are), ``illusiveness of preparedness''  (e.g. fatalistic attitudes and denial), beliefs about who is responsible for mitigating risks (e.g. individuals or governments) and how prepared one feels \citep{yong2017risk,yong2019getting,prati2011social}.

People are less likely to comply when they are facing loss of income, personal logistical problems (e.g. how to get groceries), isolation, and psychological stress (e.g. fear, boredom, frustration, stigma) \citep{digiovanni2004factors}.  Barriers to following recommendations include job insecurity, lack of childcare, guilt and anxiety about work not being completed, and the personal cost of following government advice \citep{teasdale2012importance,blake2010employment}. 

For employees, experiencing negative life events such as disasters is associated with absenteeism and lower quality of workdays \citep{north2010business}. Employers therefore need work-specific strategies and support for their employees. Employers can give employees a sense of security and help them return to work by continuing to pay full salaries on time, reassuring employees they they are not going to lose their job, having flexible work demands, implementing an organized communication strategy, and ensuring access to utilities (e.g. telephone, internet, water, electricity, sanitation) and organisational resources \citep{north2010business, donnelly2015disrupted, blake2010employment}.  

Work-specific strategies and support are also needed to ensure business continuation and survival. The disruption of activities in disasters simultaneously curtails revenues and reduces productive capacity due to the ambiguity and priorities shifting in individuals, organizations and communities \citep{donnelly2015disrupted}. As social distancing closes worksites and reduces commerce, governments face increased economic pressure to end social distancing requirements prematurely \citep{loose2010economic}. Maintaining remote workers' health and productivity is therefore important for maintaining social distancing as long as is necessary \citep{blake2010employment}. 

As we prepare this article, many other studies of the COVID-19 pandemic's effects are underway. Early evidence suggests complicated effects on productivity, which vary by person, project and metric \citep{bao2020does}. Some evidence suggests programmers are working longer hours, at an unsustainable pace \citep{forsgren2020octoverse}.

\subsection{Working from home}
\citet{perez2004technology} defines \textit{teleworking} (also called \textit{remote working}) as ``organisation of work by using information and communication technologies that enable employees and managers to access their labour activities from remote locations'' (p. 280). It includes working from home, a satellite office, a telework centre or even a coffee shop.  Remote working can help restore and maintain operational capacity and essential services during and after disasters \citep{blake2010employment}, especially when workplaces are inaccessible. Indeed, many executives are already planning to shift ``at least 5\% of previously on-site employees to permanently remote positions post-COVID 19'' \citep{Lavelle2020Gartner}. 

However, many organisations lack appropriate plans, supportive policies, resources or management practices for practising home-based telework. In disasters such as pandemics where public facilities are closed and people are required to stay at home, their experience and capacity to work can be limited by lack of dedicated workspace at home, caring responsibilities and organisational resources \citep{donnelly2015disrupted}.

In general, working from home is often claimed to improve productivity \citep{davenport1998two, mcinerney1999working, cascio2000managing} and teleworkers consistently report increased perceived productivity \citep{duxbury1998telework, baruch2000teleworking}. Interestingly, \cite{baker2007satisfaction} found that organisational and job-related factors (e.g. management culture, human resources support, structure of feedback) are more likely to affect teleworking employees' satisfaction and perceived productivity than work styles (e.g. planning vs. improvising) and household characteristics (e.g. number of children). While increasing productivity, ``working from home is associated with greater levels of both work pressure and work–life conflict'' \citep{russell2009impact} because work intrudes into developers' home lives through working unpaid overtime, thinking about work in off hours, exhaustion and sleeplessness \citep{hyman2003work}. 

Moreover, individuals' wellbeing while working remotely is influenced by their emotional stability (that is, a person's ability to their control emotions when stressed). For people with high emotional stability, working from home provides more autonomy and fosters wellbeing; however, for employees with low emotional stability, it can exacerbate physical, social and psychological strain \citep{perry2018stress}. The COVID-19 pandemic has not been good for emotional stability \citep{ARI2020worry}. 

Research on working from home has been criticized for relying on self-reported perceived productivity, which may inflate the its benefits \citep{bailey2002review}; however, objective measures often lack construct validity \citep{ralph2018construct} and perceived productivity correlates well with managers' appraisals \citep{baruch1996self}. (The perceived productivity scale we use below correlates well with objective performance data; cf. Section \ref{sec:instrument-design}).  

\subsection{Productivity and Wellbeing}

Previous studies suggest that productivity affects project outcomes and is affected by numerous factors including team size and technologies used \citep{mcleod2011factors}. However, existing research on developer productivity is rife with construct validity problems. 

Productivity is the amount of work done per unit of time. Measuring time is simple but quantifying the work done by a software developer is not. Some researchers \citep[e.g.][]{DBLP:books/sp/19/JaspanS19} argue for using goal-specific metrics. Others reject the whole idea of measuring productivity~\citep[e.g.][]{DBLP:books/sp/19/Ko19} not least because people tend to optimize for whatever metric is being used---a phenomenon known as \emph{Goodhart's law}~\citep{Goodhart1984, ChrystalMizen2003}.

Furthermore, simple productivity measures such as counting commits or modified lines of code in a certain period suffer from low construct validity \citep{ralph2018construct}.
The importance and difficulty of a commit does not necessarily correlate with its size.
Similarly, some developers might prefer dense, one-line solutions while others like to arrange their contributions in several lines.
Nevertheless, large companies including Microsoft still use controversial metrics such as number of pull requests as a ``proxy for productivity'' \citep{Spataro2020Helping}, and individual developers also use them to monitor their own performance~\citep{DBLP:conf/sigsoft/Baltes018}.
Copious time tracking tools exist for that purpose---some specifically tailored for software developers.\footnote{e.g. \url{https://wakatime.com/}}

While researchers have adapted existing scales to measure related phenomena like happiness~\citep[e.g.][]{DBLP:books/sp/19/GraziotinF19}, there is no widespread consensus about how to measure developers' productivity or the main antecedents thereof. 
Many researchers use simple, unvalidated productivity scales; for example, \cite{DBLP:journals/pacmhci/MeyerM0F17} used a single question asking participants to rate themselves from ``not productive'' to ``very productive.'' (The perceived productivity scale we use below has been repeatedly validated in multiple domains; cf. Section \ref{sec:instrument-design}).

Meanwhile, a programmer's productivity is closely related to their job satisfaction \citep{storey2019towards} and emotional state \citep{wrobel2013emotions, graziotin2015you}. Unhappiness, specifically, leads to ``low cognitive performance, mental unease or disorder, and low motivation'' \citep[p. 44]{graziotin2017consequences}. However, little is known about the antecedents or consequences of software professionals' physical or mental wellbeing in general.

\section{Hypotheses}
\label{sec:theory}

The related work discussed above suggests numerous hypotheses. Here we hypothesize about ``developers'' even though our survey was open to all software professionals because most respondents were developers (see Section \ref{sec:demographics}). These hypotheses were generated contemporaneously with questionnaire design---before data collection began.

\newcommand{\hypothesisOne}{Developers will have lower wellbeing while working from home due to COVID-19}
\paragraph{Hypothesis H1: \hypothesisOne.}

Stress, isolation, travel restrictions, business closures and the absence of educational, child care and fitness facilities all take a toll on those working from home. Indeed, a pandemic's severity and the uncertainty and isolation it induces create frustration, anxiety and fear \citep{taha2014intolerance,digiovanni2004factors,teasdale2012importance}. It therefore seems likely that many developers will be experiencing reduced emotional wellbeing. 

\newcommand{\hypothesisTwo}{Developers will have lower perceived productivity while working from home due to COVID-19}
\paragraph{Hypothesis H2: \hypothesisTwo.}

Similarly, stress, moving to an impromptu home office, and lack of child care and other amenities may have a negative impact on many developers' productivity. Many people are likely more distracted by the people they live with and their own worrisome thoughts. People tend to experience lower motivation, productivity and commitment while working from home in a disaster situation \citep{donnelly2015disrupted}. 

\begin{figure}
\includegraphics[width=0.7\textwidth]{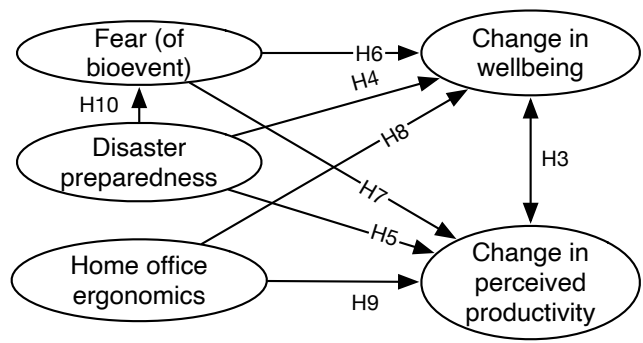}
\caption{Theoretical model of developer wellbeing and productivity}
\label{fig:theory} 
\end{figure}

\bigskip \noindent Assuming Hypotheses H1 and H2 are supported, we want to propose a model that explains and predicts changes in wellbeing and productivity (Figure \ref{fig:theory}). Hypotheses H1 and H2 are encapsulated in the \textit{change in wellbeing} and \textit{change in perceived productivity} constructs. The model only makes sense if wellbeing and productivity have changed since developers began working from home.

\newcommand{\hypothesisThree}{Change in wellbeing and change in perceived productivity are directly related}
\paragraph{Hypothesis H3: \hypothesisThree.}

We expect wellbeing and productivity to exhibit reciprocal causality. That is, as we feel worse, we become less productive, and feeling less productive makes us feel even worse, in a downward spiral. Many studies show that productivity and wellbeing covary \citep[cf.][]{dall2016characteristics}. Moreover, \cite{evers2014examining} found that people with increasing health risks have a lower wellbeing and higher dissatisfaction in life, leading to higher rates of depression and anxiety. On the other hand, decreasing health risk will increase physical and emotional wellbeing and productivity.

\paragraph{Hypotheses H4 and H5: Disaster preparedness is directly related to change in wellbeing and change in perceived productivity.}

Disaster preparedness is the degree to which a person is ready for a natural disaster. It includes behaviors like having an emergency supply kit and complying with directions from authorities. We expect lack of preparedness for disasters in general and for COVID-19 in particular to exacerbate reductions in wellbeing and productivity, and vice versa \citep[cf.][]{paton2008community,donnelly2015disrupted}.   

\paragraph{Hypotheses H6 and H7: Fear (of the pandemic) is inversely related to change in wellbeing and change in perceived productivity.}

Fear is a common reaction to bioevents like pandemics. Emerging research on COVID-19 is already showing a negative effect on wellbeing, particularly anxiety \citep{harper2020functional,xiang2020covid}. Meanwhile, fear of infection and public health measures cause psychosocial distress, increased absenteeism and reduced productivity \citep{shultz2016role,thommes2016absenteeism}. 

\paragraph{Hypotheses H8 and H9: Home office ergonomics is directly related to change in wellbeing and change in perceived productivity.}

Here we use ergonomics in its broadest sense of the degree to which an environment is safe, comfortable and conducive to the tasks being completed in it. We are not interested in measuring the angle of a developer's knees and elbows, but in a more general sense of their comfort. Professionals with more ergonomic home offices should have greater wellbeing and be more productive. \cite{donnelly2015disrupted} found that availability of a dedicated work-space at home, living circumstances, and the availability of organisational resources to work relate to the capacity to return to work after a disaster and employees' productivity. 

\paragraph{Hypothesis H10: Disaster preparedness is inversely related to fear (of the pandemic).}

It seems intuitive that the more prepared we are for a disaster, the more resilient and less afraid we will be when the disaster occurs. Indeed, Ronan et al.'s \citeyear{ronan2015disaster} systematic review found that programs for increasing disaster preparedness had a small- to medium-sized negative effect on fear. People who have high self-efficacy and response-efficacy (i.e. perceive themselves as ready to face a disaster) will be less afraid \citep{roberto2009raising}.  

\section{Method}
\label{sec:method}

On March 19, 2020, the first author initiated a survey to investigate how COVID-19 affects developers, and recruited the second and third authors for help. We created the questionnaire and it was approved by Dalhousie University's research ethics board in less than 24 hours. We began data collection on March 27th. We then recruited authors 5 through 17, who translated and localized the questionnaire into \languages, and created region-specific advertising strategies. Translations launched between April 5 and 7, and we completed data collection between April 12 and 16. Next, we recruited the fourth author to assist with the data analysis, which was completed on April 29. The manuscript was prepared primarily by the first four authors with edits from the rest of team. 

This section details our approach and instrumentation. 

\subsection{Population and inclusion criteria}

This study's target population is software developers anywhere in the world \textit{who switched from working in an office to working from home} because of COVID-19. Of course, developers who had been working remotely before the pandemic and developers who continued working in offices throughout the pandemic are also important, but this study is about the switch, and the questions are designed for people who switched from working on-site to at home. 

In principle, the questionnaire was open to all sorts of software professionals, including designers, quality assurance specialists, product managers, architects and business analysis, but we are mainly interested in developers, our marketing focuses on software developers, and therefore most respondents identify as developers (see Section \ref{sec:demographics}). 

\subsection{Instrument design}
\label{sec:instrument-design}

We created an anonymous questionnaire survey. We did not use URL tracking or tokens. We did not collect contact information. 

Questions were organized into blocks corresponding to scale or question type. The order of the items in each multi-item scale was randomized to mitigate primacy and recency effects. The order of blocks was not randomized because our pilot study (Section \ref{sec:pilot}) suggested that the questionnaire was more clear when the questions that distinguish between before and after the switch to home working came after those that do not. 

The questionnaire used a filter question to exclude respondents who do not meet the inclusion criteria. Respondents who had not switched from working in an office to working from home because of COVID-19 simply skipped to the end of the questionnaire. It also included not only traditional demographic variables (e.g. age, gender, country, experience, education) but also how many adults and children (under twelve) participants lived with, the extent to which participants are staying home and whether they or any friends or family had tested positive for COVID-19. 

The questionnaire used validated scales as much as possible to improve construct validity.  A \textit{construct} is a quantity that cannot be measured directly. Fear, disaster preparedness, home office ergonomics, wellbeing and productivity are all constructs. In contrast, age, country, and number of children are all directly measurable. Direct measurements are assumed to have inherent validity, but latent variables have to be validated to ensure that they measure the right properties \citep[cf.][]{ralph2018construct}.

The exact question wording can be seen in our replication pack (see Section \ref{sec:dataAvailability}). This section describes the scales and additional questions.  
\paragraph{Emotional wellbeing (WHO-5).}

To assess emotional wellbeing, we used the WHO's five-item wellbeing index (WHO-5).\footnote{\url{https://www.psykiatri-regionh.dk/who-5/Documents/WHO5_English.pdf}} Each item is assessed on a six-point scale from ``at no time'' (0) to ``all of the time'' (5). The scale can be calculated by summing the items or using factor analysis. The WHO-5 scale is widely used, widely applicable, and has high sensitivity and construct validity \citep{topp20155}. Respondents self-assessed their wellbeing twice: once for the four weeks prior to beginning to work from home, and then again for the time they have been working from home.  

\paragraph{Perceived Productivity (HPQ).}

To assess perceived productivity we used items from the WHO's Health and Work Performance Questionnaire (HPQ).\footnote{\url{https://www.hcp.med.harvard.edu/hpq/info.php}} The HPQ measures perceived productivity in two ways: (1) using an eight-item summative scale, with multiple reversed indicators, that assesses overall and relative performance; and (2) using 11-point general ratings of participants' own performance and typical performance of similar workers. These scales are amenable to factor analysis or summation. Of course, people tend to overestimate their performance relative to their peers, but we are comparing participants to their past selves not to each other. HPQ scores are closely related to objective performance data in diverse fields \citep{kessler2003world}. Again, respondents self-assessed their productivity for both the four weeks prior to working from home, and for the time they have been working from home.   

\paragraph{Disaster Preparedness (DP).} 

To assess disaster preparedness, we adapted Yong et al.'s (\citeyear{yong2017risk}) individual disaster preparedness scale. Yong et al.\ developed their five-item, five-point, Likert scale based on common, important behaviors such as complying with government recommendations and having emergency supplies. The scale was validated using a questionnaire survey of a ``weighted nationally representative sample'' of 1084 Canadians. We adapted the items to refer specifically to COVID-19. It can be computed by summing the responses or using factor analysis.

\paragraph{Fear and Resilience (FR).}

The Bracha-Burkle Fear and Resilience (FR) checklist is a triage tool for assessing patients' reactions to bioevents (including pandemics). The FR checklist places the patient on a scale from intense fear to hyper-resilience \citep{bracha2006utility}. We dropped some of the more extreme items (e.g. ``Right now are you experiencing shortness of breath?'') because respondents are at home taking a survey, not arriving in a hospital emergency room. The FR checklist is a weighted summative scale so it has to be computed manually using Bracha and Burkle's formula rather than using factor analysis. It has multiple reversed indicators.

\paragraph{Ergonomics.}

We could not find a reasonable scale for evaluating home office ergonomics. There is comparatively less research on the ergonomics of home offices \citep{inalhan2010teleworker} and ergonomic instruments tend to be too narrow (e.g. evaluating hip angle). Based on our reading of the ergonomics literature, we made a simple six-item, six-point Likert scale concerning distractions, noise, lighting, temperature, chair comfort and overall ergonomics. 

Again, we evaluated the scale's face and content validity using a pilot study (see Section~\ref{sec:pilot}) and examine convergent and discriminant validity ex post in Section~\ref{sec:validityAnalysis}.

\paragraph{Organizational Support (OS).} 

We could not find any existing instrument that measures the degree to which an organization supports its employees during a crisis. The first author therefore interviewed three developers with experience in both co-located and distributed teams as well as office work and working from home. Interviewees brainstormed actions companies could take to help, and we used open-coding \citep{saldana2015coding} to organize their responses into five themes: 

\begin{enumerate}
    \item \textit{Equipment:} providing equipment employees need in their home office (e.g. a second monitor) 
    \item  \textit{Reassurance:} adopting a tone that removes doubt and fear (e.g. assuring employees that lower productivity would be understood)  
    \item \textit{Connectedness:} encouraging virtual socializing (e.g. through video chat)
    \item \textit{Self-care:} providing personal services not directly related to work (e.g. resources for exercising or home-schooling children)   
    \item \textit{Technical infrastructure and practices:} ensuring that remote infrastructure (e.g. VPNs) and practices (e.g. code review) are in place.  
\end{enumerate}

We generated a list of 22 actions (four or five per theme) by synthesizing the ideas of interviewees with existing literature on working from home, distributed development and software engineering more generally. For each action, respondents indicate whether their employer is taking the action and whether they think it is or would be helpful. Organizational support is not a construct in our theory per se because we have insufficient a priori information to produce a quantitative estimate, so we analyze these answers separately.  

\subsection{Pilot}
\label{sec:pilot}

We solicited feedback from twelve colleagues: six software engineering academics and six experienced software developers. Pilot participants made various comments on the questionnaire structure, directions and on the face and content validity of the scales. Based on this feedback we made numerous changes including clarifying directions, making the question order static, moving the WHO-5 and HPQ scales closer to the end, dropping some problematic questions, splitting up an overloaded question, and adding some open response questions. (Free-text answers are not analyzed in this paper; open response questions were included mainly to inform future research; see Section~\ref{sec:futureDirections}). 

\subsection{Sampling, localization and incentives}

We advertised our survey on social and conventional media, including \textsf{Dev.to, Développez.com, DNU.nl, eksisozluk, Facebook, Hacker News, Heise Online, \mbox{InfoQ,} LinkedIn, Twitter, Reddit} and \textsf{WeChat}. Upon completion, participants were provided a link and encouraged to share it with colleagues who might also like to take the survey. Because this is an anonymous survey, we did not ask respondents to provide colleagues' email addresses. 

We considered several alternatives, including scraping emails from software repositories and stratified random sampling using company lists, but none of these options seemed likely to produce a more representative sample. Granted, if we sampled from an understood sampling frame, we could better evaluate the representativeness of the sample and generalizability of the results; however, we are not aware of any sampling frames with sufficiently well-understood demographics to facilitate accurate inferences. 

Instead, we focused on increasing the diversity of the sample by localizing the survey and promoting it in more jurisdictions. We translated the survey into \languages. Each author-translator translated from English into their first language. We capitalized on each authors' local knowledge to reach the more people in their jurisdiction. Rather than a single, global campaign, we used a collection of local campaigns.  

Each localization involved small changes in wording. Only a few significant changes were needed. The Chinese version used a different questionnaire system (\url{wjx.cn}) because Google Forms is not available in China. Furthermore, because the lockdowns in China and Korea were ending, we reworded some questions from ``since you began working from home'' to ``while you were working from home.''


We did not offer cash incentives for participation. Rather, we offered to donate US\$500 to an open source project chosen by participants (in one of the open response questions). Respondents suggested a wide variety of projects, so we donated US\$100 to the five most mentioned: The Linux Foundation, The Wikimedia Foundation, The Mozilla Foundation, The Apache Software Foundation and the Free Software Foundation. The Portuguese version was slightly different: it promised to donate 1000 BRL to Ação da Cidadania's (a Brazillian NGO) Action against Corona project rather than a project chosen by participants (which we did).  

\section{Analysis and Results}
\label{sec:results}
We received \totalResponses total responses of which \invalidResponses did not meet our inclusion criteria and \blankResponses were effectively blank (see below) leaving \validResponses. 
This section describes how the data was cleaned and analyzed.

\subsection{Data cleaning}

The data was cleaned as follows. 
\begin{enumerate}
    \item Delete responses that do not meet inclusion criteria. 
    \item Delete almost empty rows, where the respondent apparently answered the filter question correctly, then skipped all other questions.
    \item Delete the timestamp field (to preserve anonymity), the consent form confirmation field (because participants could not continue without checking these boxes so the answer is always ``TRUE'') and the filter question field (because all remaining rows have the same answer). 
    \item Add a binary field indicating whether the respondent had entered text in at least one long-answer question (see Section~\ref{sec:validityAnalysis})
    \item Move all free-text responses to a separate file (to preserve anonymity). 
    \item Recode the raw data (which is in different languages with different alphabets) into a common quantitative coding scheme; for example, from 1 for ``strongly disagree'' to 5 for ``strongly agree'' The recoding instructions and related scripts are included in our replication package (see Section \ref{sec:dataAvailability}). 
    \item Split select-multiple questions into one binary variable per checkbox (Goo\-gle Forms uses a comma-separated list of the text of selected answers).  
    \item Add a field indicating the language of the response.
    \item Combine the responses into a single data file. 
    \item Calculate the FR scale according to its formula \citep{bracha2006utility}.
\end{enumerate}

\subsection{Validity analysis}
\label{sec:validityAnalysis}

We evaluated construct validity using established guidelines~\citep{ralph2018construct}. First, we assessed content validity using a pilot study (Section~\ref{sec:pilot}). Next, we assessed convergent and discriminant validity using a principle component analysis (PCA) with Varimax rotation and Kaiser normalization. Bart\-lett's test is significant ($chi-square=13229; df=253; p<0.001$) and our KMO measure of sampling adequacy is high (0.874), indicating that our data is appropriate for factor analysis. 

\begin{table}
\caption{First principle components analysis*}
\label{tab:PCA1}

\begin{tabular}{lcccc}
\hline\noalign{\smallskip}
Variable & \multicolumn{4}{c}{ Component } \\
& 1 & 2 & 3 & 4 \\
\noalign{\smallskip}\hline\noalign{\smallskip}
$\Delta$ P8 & 0.740 & & & \\
$\Delta$ P2 & 0.715 & & & \\
$\Delta$ P9 & 0.704 & \textbf{0.304} & & \\
$\Delta$ P6 & 0.699 & & & \\
$\Delta$ P4 & 0.669 & & & \\
$\Delta$ P3 & 0.645 & & & \\
$\Delta$ P5 & 0.64 & & & \\
$\Delta$ P1 & 0.563 & & & \\
$\Delta$ P7 & \textbf{0.356} & & & \\
$\Delta$ WP1 & & 0.838 & & \\
$\Delta$ WP2 & & 0.791 & & \\
$\Delta$ WP3 & & 0.782 & & \\
$\Delta$ WP5 & & 0.734 & & \\
$\Delta$ WP4 & & 0.727 & & \\
Erg6 & & & 0.802 & \\
Erg5 & & & 0.748 & \\
Erg2 & & & 0.666 & \\
Erg3 & & & 0.645 & \\
Erg1 & \textbf{0.306} & & 0.640 & \\
Erg4 & & & 0.628 & \\
DP3 & & & & 0.688 \\
DP1 & & & & 0.661 \\
DP5 & & & & 0.568 \\
DP2 & & & & 0.565 \\
DP4 & & & & 0.493 \\
\noalign{\smallskip}\hline
\end{tabular}

*Rotation converged in 5 iterations.
Coefficients $<0.3$ suppressed.

\end{table}

As Table~\ref{tab:PCA1} shows, the items load well but not perfectly. The bold coefficients suggest possible issues with Change in Productivity ($\Delta$ Productivity) 7 and 9, as well as Ergonomics 1. We dropped items one at a time until the loadings stabilized, starting with $\Delta$ Productivity 7, followed by $\Delta$ Productivity 9. As shown in Table~\ref{tab:PCA2}, dropping these two indicators solved the problem with Ergonomics 1, so the latter is retained. 

\begin{table}
\caption{Second principle components analysis}
\label{tab:PCA2}

\begin{tabular}{lcccc}
\hline\noalign{\smallskip}
Variable & \multicolumn{4}{c}{ Component } \\
& 1 & 2 & 3 & 4 \\
\noalign{\smallskip}\hline\noalign{\smallskip}
$\Delta$ P2 & 0.721 & & & \\
$\Delta$ P8 & 0.718 & & & \\
$\Delta$ P6 & 0.703 & & & \\
$\Delta$ P4 & 0.679 & & & \\
$\Delta$ P3 & 0.651 & & & \\
$\Delta$ P5 & 0.649 & & & \\
$\Delta$ P1 & 0.566 & & & \\
$\Delta$ WB1 & & 0.845 & & \\
$\Delta$ WB2 & & 0.797 & & \\
$\Delta$ WB3 & & 0.790 & & \\
$\Delta$ WB5 & & 0.740 & & \\
$\Delta$ WB4 & & 0.732 & & \\
Erg6 & & & 0.803 & \\
Erg5 & & & 0.745 & \\
Erg2 & & & 0.669 & \\
Erg1 & & & 0.646 & \\
Erg3 & & & 0.644 & \\
Erg4 & & & 0.629 & \\
DP3 & & & & 0.685 \\
DP1 & & & & 0.666 \\
DP2 & & & & 0.570 \\
DP5 & & & & 0.565 \\
DP4 & & & & 0.490 \\
\noalign{\smallskip}\hline
\end{tabular}

\emph{Notes: Rotation convErged in 5 iterations; correlations $<0.3$ suppressed.}

\end{table}

We evaluate predictive validity by testing our hypotheses in Section \ref{sec:hypothesisTesting}.

\paragraph{Response bias.}

Here, \textit{response bias} refers to the possibility that people for whom one of our hypotheses hold are more likely to take the questionnaire, thus inflating the results.

There are two basic ways to analyze this kind of response bias. The first---comparing sample parameters to known population parameters---is impractical because no one has ever established population parameters for software professionals. The second---comparing late respondents to early respondents---cannot be used because we do not know the time between each respondent learning of the survey and completing it. However, we can do something similar: we can compare respondents who answered one or more open response questions (more keen on the survey) with those who skipped the open response questions (less keen on the survey). 

As shown in Table \ref{tab:responseBias}, only number of adult cohabitants and age have significant differences, and in both cases, the effect size ($\eta^2$) is very small. This is consistent with minimal response bias. 

\begin{table}
\caption{Analysis of response bias (one-way ANOVA)}
\label{tab:responseBias}

\begin{tabular}{lrrr}
\hline\noalign{\smallskip}
Variable & F & Sig. & $\eta^2$ \\
\noalign{\smallskip}\hline\noalign{\smallskip}
age & 4.250 & 0.039 & 0.002 \\
disability & 0.117 & 0.733 & 0.000 \\
education & 0.153 & 0.696 & 0.000 \\
adultCohabitants & 19.037 & 0.000 & 0.009 \\
childCohabitants & 0.358 & 0.550 & 0.000 \\
experience & 3.381 & 0.066 & 0.002 \\
remoteExperience & 0.013 & 0.910 & 0.000 \\
organizationSize & 0.330 & 0.566 & 0.000 \\
\noalign{\smallskip}\hline
\end{tabular}

\end{table}

\subsection{Demographics}
\label{sec:demographics}

\begin{figure}
\centering
\includegraphics[width=0.9\columnwidth, trim=0.0in 0.6in 0.6in 0.3in]{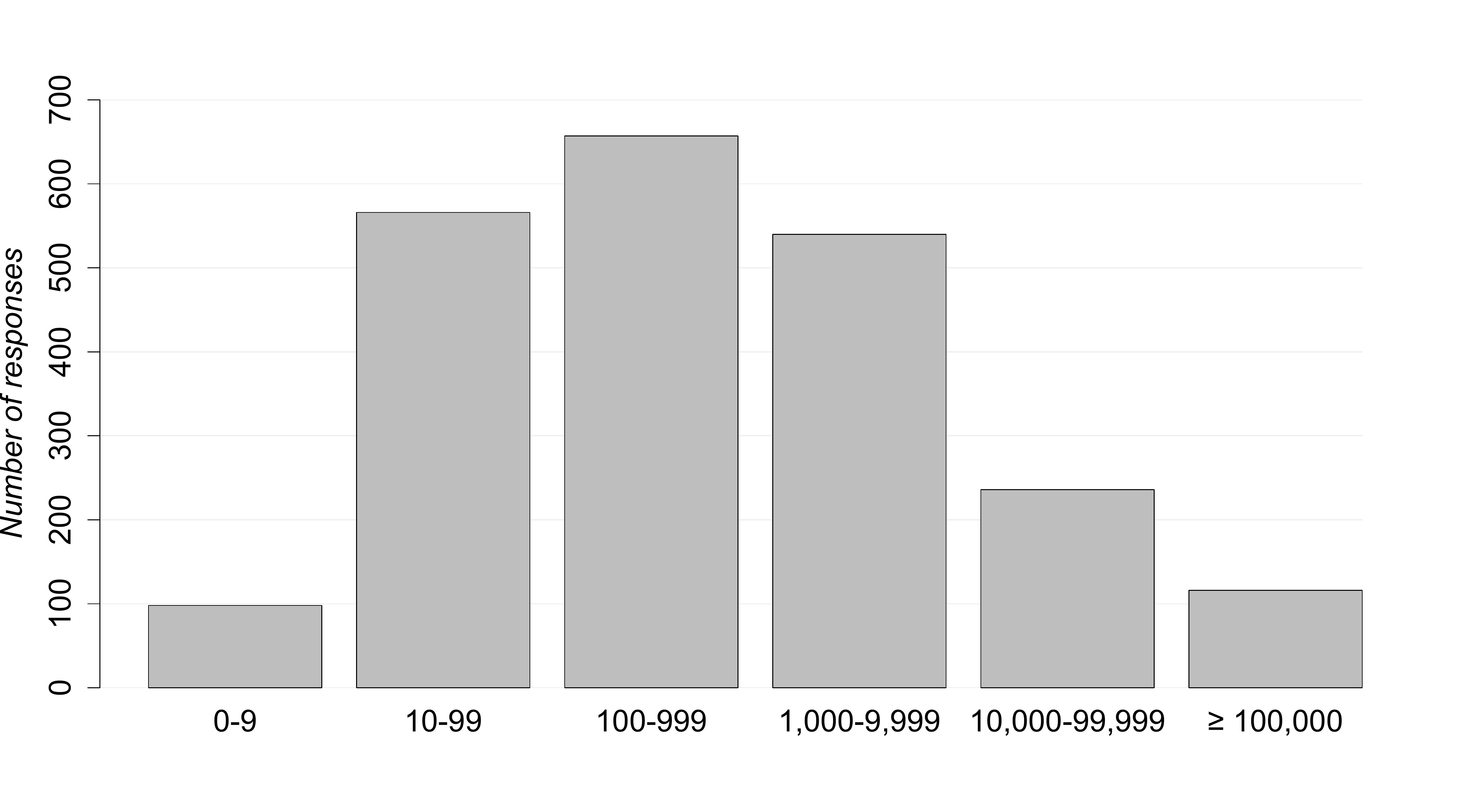} 
\caption{Organization sizes}
\label{fig:orgsize} 
\end{figure}

\begin{figure}
\centering
\includegraphics[width=0.9\columnwidth, trim=0.0in 0.6in 0.6in 0.3in]{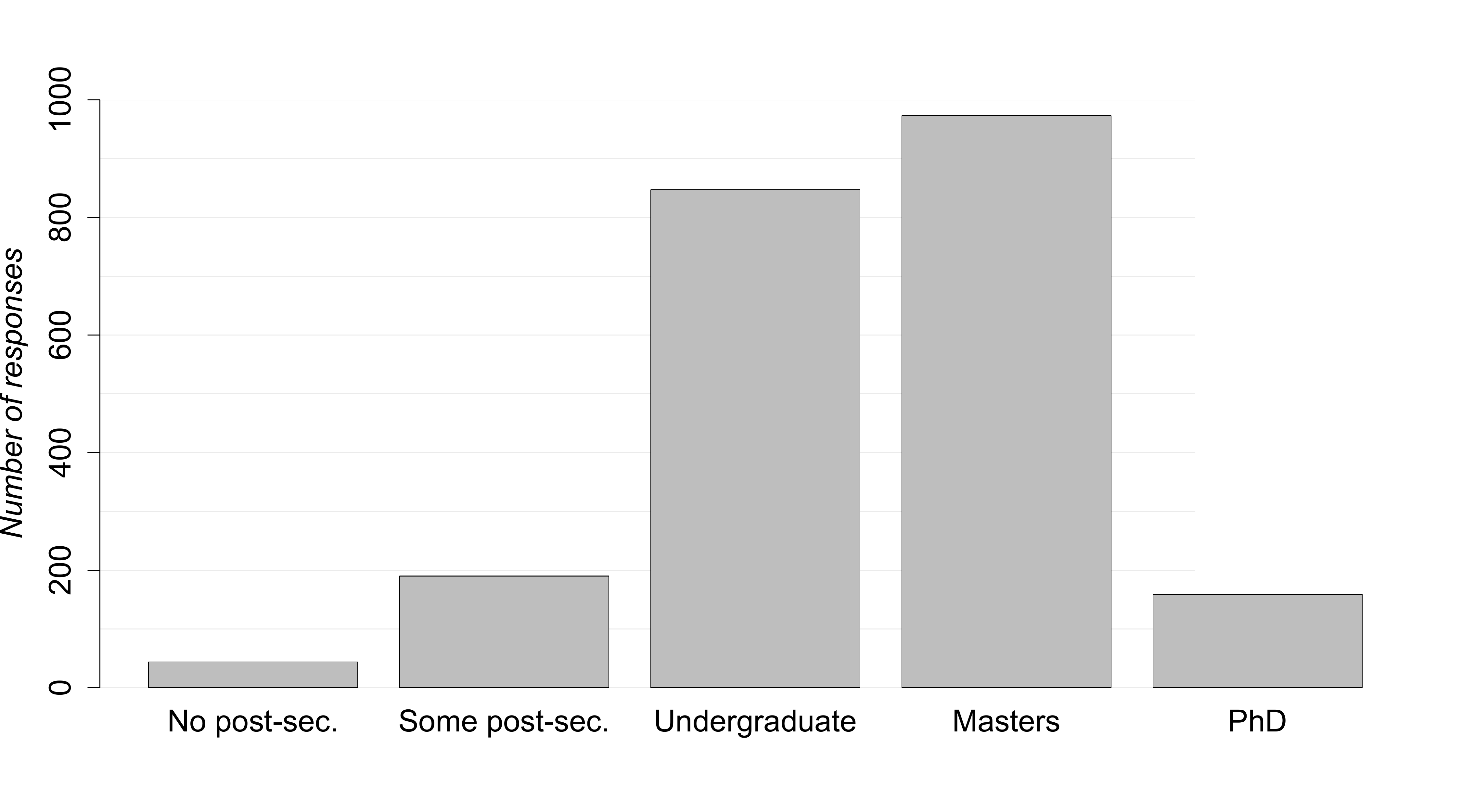} 
\caption{Participants' education levels}
\label{fig:education} 
\end{figure}

Respondents were disproportionately male (81\% vs.\ 18\% female and 1\% non-binary) 
and overwhelmingly employed full-time (94\%) 
with a median age of 30--34. 
Participants were generally well-educated (Fig.~\ref{fig:education}).
Most respondents (53\%) live with one other adult, while 18\% live with no other adults and the rest live with two or more people. 
27\% live with one or more children under 12. 
8\% indicate that they may have a disability that affects their work. 
Mean work experience is 9.3 years ($\sigma = 7.3$). 
Mean experience working from home is 1.3 years ($\sigma = 2.5$); however, 58\% of respondents have no experience working from home.  

Participants hail from \numberOfCountries countries (Table \ref{tab:countries}) and organizations ranging from 0--9 employees to more than 100,000 (Fig~\ref{fig:orgsize}). Many participants listed multiple roles but 80\% included software developer or equivalent among them, while the rest were other kinds of software professionals (e.g. project manager, quality assurance analyst). 

\begin{table}
\caption{Respondents' countries of residence}
\label{tab:countries}

\begin{tabular}{lrr|lrr}
\hline\noalign{\smallskip}
Country & n & \% & Country & n & \% \\
\noalign{\smallskip}\hline\noalign{\smallskip}
Germany & 505 & 22.7\% & Japan & 53 & 2.4\% \\
Russia & 366 & 16.4\% & Spain & 52 & 2.3\% \\ 
Brazil & 272 & 12.2\% & Iran & 40 & 1.8\% \\
Italy & 173 & 7.8\% & Austria & 29 & 1.3\% \\
United States & 99 & 4.4\% & Canada & 27 & 1.2\% \\
South Korea & 81 & 3.6\% & Switzerland & 20 & 0.9\% \\
Belgium & 77 & 3.5\% & United Kingdom & 20 & 0.9\% \\
China & 76 & 3.4\% & n/a & 20 & 0.9\% \\
Turkey & 66 & 3.0\% & Other & 194 & 8.7\% \\
India & 55 & 2.5\% \\
\noalign{\smallskip}\hline
\end{tabular}

\end{table}

Seven participants ($<1\%$) tested positive for COVID-19 and six more ($<1\%$) live with someone with COVID-19; 4\% of respondents indicated that a close friend or family member had tested positive, and 13\% were currently or recently quarantined.

\subsection{Change in wellbeing and productivity}
\label{sec:hypothesisTesting}

\paragraph{Hypothesis H1: \hypothesisOne.}

\begin{figure}
\includegraphics[width=\textwidth]{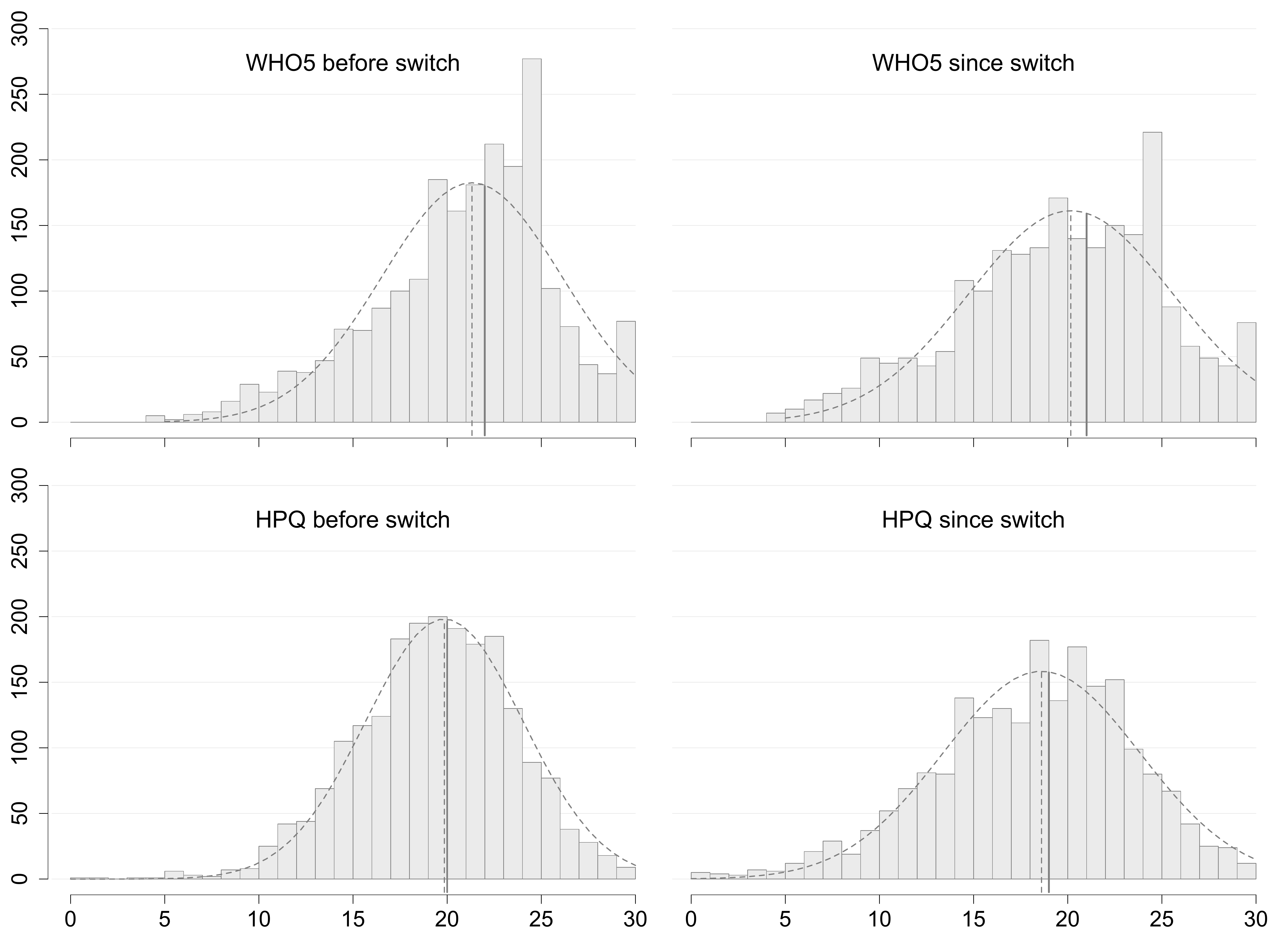}
\caption{Distribution of ratings on the WHO5 and HPQ scales before and since switching to working form home with mean (dashed line) and median (solid line) values (2,194 complete cases for WHO5 and 2,078 for HPQ)}
\label{fig:WHO5-HPQ}
\end{figure}

Participants responded to the WHO5 wellbeing scale twice---once referring to the 28-day period before switching to work from home and once referring to the period while working from home. We estimate wellbeing before and after by summing each set of five items, and then compare the resulting distributions (see Figure~\ref{fig:WHO5-HPQ}). Since both scales deviate significantly from a normal distribution (Shapiro-Wilk test; $p<0.001$; see Fig. \ref{fig:WHO5-HPQ}), we compare the distributions using the two-sided paired Wilcoxon signed rank test. To estimate effect size, we use Cliff's delta (with 95\% confidence level). 

\textbf{Hypothesis H1 is supported} (Wilcoxon signed rank test $V=645610$; $p<0.001$; $\delta=0.12 \pm 0.03$). 

\paragraph{Hypothesis H2: \hypothesisTwo.}

Like the wellbeing scale, participants answered the HPQ productivity scale twice. Again, we estimate productivity before and after by summing each set set of items (after correcting reversed items and omitting items 7 and 9; see Section~\ref{sec:validityAnalysis}). Again, the distributions are not normal (Shapiro-Wilk test; $p<0.001$; see Fig. \ref{fig:WHO5-HPQ}), so we use the Wilcoxon signed rank test and Cliff's delta. 

\textbf{Hypothesis H2 is supported} (Wilcoxon signed rank test $V=566520$; $p<0.001$; $\delta=0.13 \pm 0.03$).

\subsection{Structural equation model}
\label{sec:SEM}

To test our remaining hypotheses, we use structural equation modeling (SEM). Briefly, SEM is used to test theories involving constructs (also called latent variables). A \textit{construct} is a quantity that cannot be measured directly~\citep{ralph2018construct}. Fear, disaster preparedness, home office ergonomics, wellbeing and productivity are all constructs. In contrast, age, country, and number of children are all directly measurable. 

To design a structural equation model, we first define a \textit{measurement model}, which maps each \textit{reflective indicator} into its corresponding construct. For example, each of the five items comprising the WHO5 wellbeing scale is modeled as a reflective indicator of wellbeing. SEM uses confirmatory factor analysis to estimate each construct as the shared variance of its respective indicators. 

Next, we define the \textit{structural model}, which identifies the expected relationships among the constructs. The constructs we are attempting to predict are referred to as \textit{endogenous}, while the predictors are \textit{exogenous}. 

SEM uses a path modeling technique (e.g. regression) to build a model that predicts the endogenous (latent) variables based on the exogenous variables, and to estimate both the strength of each relationship and the overall accuracy of the model.\footnote{Data was analyzed using the \textsf{R} package \textsf{lavaan} 0.6-5. available at \url{http://lavaan.ugent.be/}.}

As mentioned, the first step in a SEM analysis is to conduct a confirmatory factor analysis to verify that the measurement model is consistent (Table \ref{tab:CFA}). Here, the latent concepts Ergonomics and DisasterPreparedness are captured by their respective exogenous variables. Fear is not included because it is computed manually (see Section \ref{sec:instrument-design}). $\Delta \mr{Wellbeing}$ is the difference in a participant's emotional wellbeing before and after switching to working from home. This latent concept is captured by five exogenous variables, $\Delta \mr{WB1},\dots, \Delta \mr{WB5}$. Similarly, $\Delta$Productivity represents the difference in perceived productivity, before and after switching to working from home.

\begin{table}
\caption{Confirmatory factor analysis}
\label{tab:CFA}

\begin{tabular}{llrrrr}
\hline\noalign{\smallskip}
Construct & Indicator & Estimate & Std.Err & z-value & $P(>|z|)$ \\
\noalign{\smallskip}\hline\noalign{\smallskip}
$\Delta \mr{Wellbeing}=\sim$ & $\Delta$WB1 & 1.000 & & & \\
& $\Delta$WB2 & 0.896 & 0.016 & 54.518 & 0 \\
& $\Delta$WB3 & 0.955 & 0.016 & 58.917 & 0 \\
& $\Delta$WB4 & 0.804 & 0.018 & 44.686 & 0 \\
& $\Delta$WB5 & 0.848 & 0.017 & 51.041 & 0 \\
& & & & & \\
$\Delta \mr{Productivity}=\sim$ & $\Delta$P1 & 1.000 & & & \\
& $\Delta$P2 & -1.268 & 0.053 & -24.084 & 0 \\
& $\Delta$P3 & -1.120 & 0.053 & -20.979 & 0 \\
& $\Delta$P4 & -1.239 & 0.053 & -23.263 & 0 \\
& $\Delta$P5 & -1.229 & 0.055 & -22.266 & 0 \\
& $\Delta$P6 & -1.306 & 0.058 & -22.677 & 0 \\
& $\Delta$P8 & 1.460 & 0.057 & 25.512 & 0 \\
& & & & & \\
$\mr{Ergonomics}=\sim$ & Erg1 & 1.000 & & & \\
& Erg2 & 0.964 & 0.035 & 27.395 & 0 \\
& Erg3 & 0.820 & 0.037 & 22.128 & 0 \\
& Erg4 & 0.937 & 0.035 & 26.663 & 0 \\
& Erg5 & 1.064 & 0.034 & 31.535 & 0 \\
& Erg6 & 1.258 & 0.035 & 35.821 & 0 \\
& & & & & \\
Disaster & DP1 & 1.000 & & & \\
$\mr{Preparedness}=\sim$ & DP2 & 0.716 & 0.089 & 8.079 & 0 \\
& DP3 & 1.181 & 0.112 & 10.521 & 0 \\
& DP4 & 0.923 & 0.105 & 8.805 & 0 \\
& DP5 & 1.186 & 0.120 & 9.888 & 0 \\
\noalign{\smallskip}\hline
\end{tabular}

\emph{Notes: converged after 50 iterations with 185 free parameters with ($n=1377$); estimates may exceed 1.0 because they are regression coefficients, not correlations as in Principal Component Analysis; negative estimates indicate reversed items}
\end{table}

The confirmatory factor analysis converged (not converging would suggest a problem with the measurement model) and all of the indicators load well on their constructs. The lowest estimate, 0.716 for DP2, is still quite good. The estimates for $\Delta$ P2 through $\Delta$ P6 are negative because these items were reversed (i.e. higher score = worse productivity). Note that factor loadings greater than one do not indicate a problem because they are regression coefficients, not correlations \citep{joreskog1999large}. 

Having reached confidence in our measurement model, we construct our structural model by representing all of the hypotheses stated in Section \ref{sec:theory} as regressions (e.g. $\Delta$Wellbeing $\sim$ DisasterPreparedness + Fear + Ergonomics). 

In principle, we use all control variables as predictors for all latent variables. In practice, however, this leads to too many relationships and prevents the model from converging. Therefore, we evaluate the predictive power of each control variable, one at a time, and include it in a regression only where it makes at least a marginally significant ($p<0.1$) difference. Here, using a higher than normal p-value is more conservative because we are dropping predictors rather than testing hypotheses. Country of residence and language of questionnaire are not included because SEM does not respond well to nominal categorical variables (see Section \ref{sec:exploratoryAnalysis}).

Since the exogenous variables are ordinal, the weighted least square mean variance (WLSMV) estimator was used. WLSMV uses diagonally weighted least squares to estimate the model parameters, but it will use the full weight matrix to compute robust standard errors, and a mean- and variance-adjusted test statistic. In short, the WLSMV is a robust estimator which does not assume a normal distribution, and provides the best option for modelling ordinal data in SEM~\citep{brown2006confirmatory}. We use the default Nonlinear Minimization subject to Box Constraints (NLMinB) optimizer. 

For missing data, we use pairwise deletion: we only keep those observations for which both values are observed (this may change from pair to pair). By default, since we are also dealing with categorical exogenous variables, the model is set to be conditional on the exogenous variables.

The model was executed and all diagnostics passed, that is, \textsf{lavaan} ended normally after 97 iterations with 212 free parameters and $n=1377$. We evaluate model fit by inspecting several indicators \citep[cf.][for cut-offs]{HuB99sem}:
\begin{itemize}
    \item The Comparative Fit Index ($\mr{CFI} = 0.961$) and Tucker-Lewis Index ($\mr{TLI} = 0.979$), which compare the model's fit against the null model, should be at least 0.95.
    \item The Root Mean Square Error of Approximation ($\mr{RMSEA} = 0.051$, 90\% CI [0.048; 0.053]) should be less than 0.06. 
    \item The Standardized Root Mean Square Residual ($\mr{SRMR}=0.067$) should be less than 0.08 (for large sample sizes).
\end{itemize}

In summary, all diagnostics indicate the model is safe to interpret (i.e. $\mr{CFI} = 0.961$, $\mr{RMSEA} = 0.051$, $\mr{SRMR} = 0.067$).

\begin{figure}
\includegraphics[width=\textwidth]{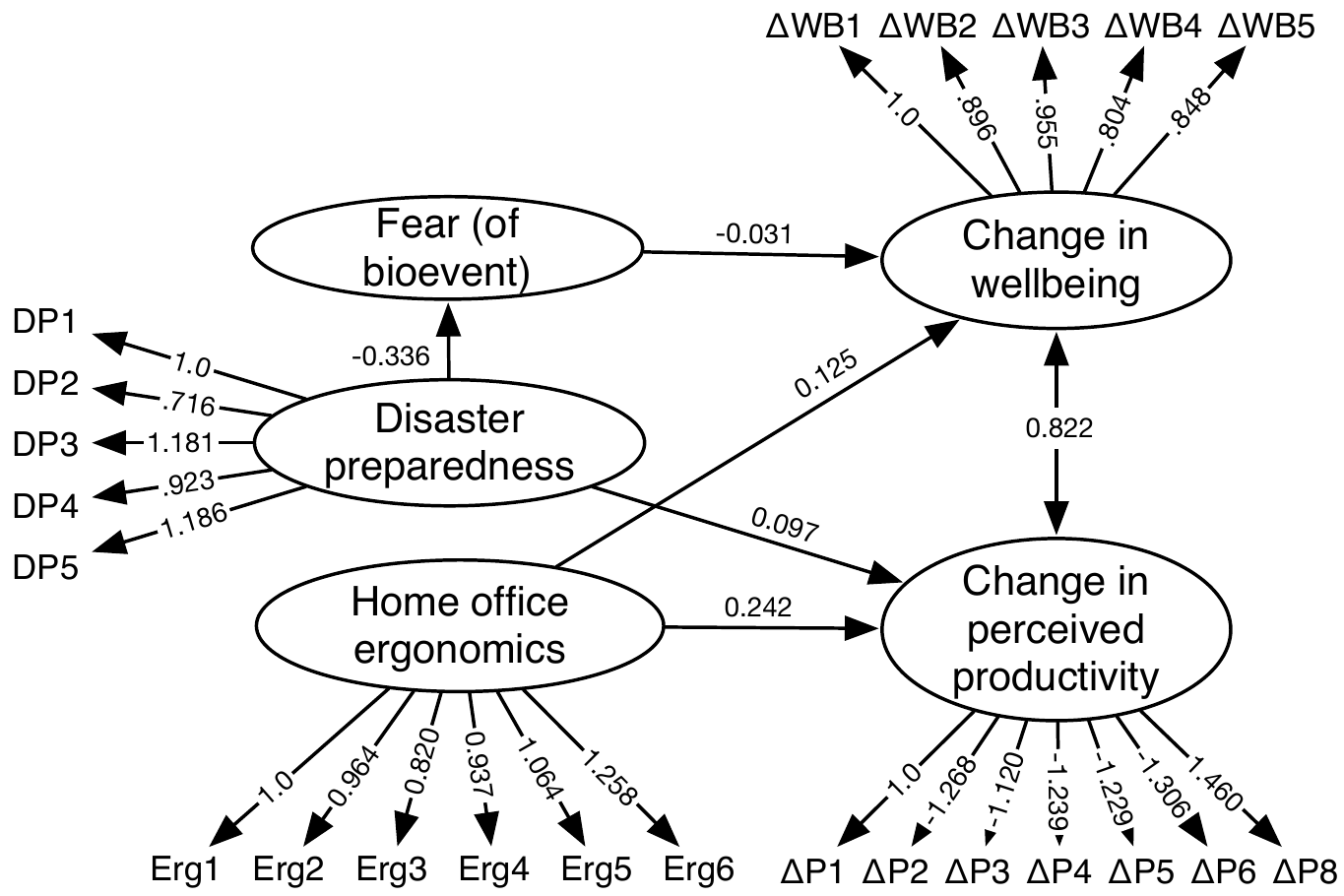}
\caption{Supported model of developer wellbeing and productivity}
\label{fig:finalModel} 
\emph{Note: error terms, unsupported hypotheses and control variables are omitted for clarity}
\end{figure}

Figure \ref{fig:finalModel} illustrates the supported structural equation model. The arrows between the constructs show the supported causal relationships. The path coefficients (the numbers on the arrows) indicate the relative strength and direction of the relationships. For example, the arrow from Disaster Preparedness to Fear indicates that the hypothesis that Disaster Preparedness affects Fear was supported. The label ($-0.336$) indicates an inverse relationship (more Disaster Preparedness leads to less Fear) and 0.336 indicates the strength of the relationship.    

\bigskip
\noindent\begin{tabular}{|p{0.95\textwidth}|}
\hline
\\
Based on this model, \textbf{Hypotheses H1--H3, H5, H6, and H8--H10 are supported; Hypotheses H4 and H7 are not supported}. That is, change in wellbeing and change in perceived productivity are directly related; change in perceived productivity depends on home office ergonomics and disaster preparedness; change in wellbeing depends on ergonomics and fear; and disaster preparedness is inversely related to fear.\\ 
\\
\hline
\end{tabular}
\bigskip

\subsection{Exploratory findings}
\label{sec:exploratoryAnalysis}

\begin{table}
\caption{Structural equation model regressions}
\label{tab:SEM}

\begin{tabular}{llrrrr}
\hline\noalign{\smallskip}
Construct & Predictor & Estimate & Std.Err & $z$-value & $P(>|z|)$ \\
\noalign{\smallskip}\hline\noalign{\smallskip}
Disaster & adultCohabitants & 0.080 & 0.019 & 4.234 & 0.000  \\
Preparedness $\sim$ & disability & -0.179 & 0.059 & -3.035 & 0.002  \\
 & covidStatus & 0.073 & 0.032 & 2.260 & 0.024 \\
 & education & -0.050 & 0.026 & -1.882 & 0.060 \\
 & & & & & \\
Ergonomics $\sim$  & children & -0.163 & 0.031 & -5.184 & 0.000 \\
 & adultCohabitants & -0.047 & 0.019 & -2.457 & 0.014 \\
 & disability & -0.110 & 0.057 & -1.932 & 0.053 \\
 & remoteExperience & 0.044 & 0.026 & 1.709 & 0.087 \\
  & & & & & \\
Fear $\sim$ & isolation & 0.502 & 0.105 & 4.764 & 0.000 \\
 & DisasterPreparedness & -0.336 & 0.106 & -3.161 & 0.002 \\
 & role & -0.356 & 0.116 & -3.056 & 0.002 \\
 & covidStatus & 0.196 & 0.075 & 2.607 & 0.009 \\
 & gender & 0.273 & 0.122 & 2.241 & 0.025 \\
 & disability & 0.265 & 0.119 & 2.227 & 0.026 \\
 & education & -0.122 & 0.060 & -2.047 & 0.041 \\
 & children & 0.116 & 0.063 & 1.831 & 0.067 \\ 
 & & & & & \\
$\Delta$Wellbeing $\sim$ & Ergonomics & 0.125 & 0.033 & 3.813 & 0.000 \\
 & covidStatus & -0.121 & 0.040 & -3.041 & 0.002  \\
 & Fear & -0.031 & 0.012 & -2.542 & 0.011 \\
 & age & 0.097 & 0.044 & 2.204 & 0.028  \\
 & DisasterPreparedness & -0.020 & 0.049 & -0.416 & 0.678 \\
 & & & & & \\
$\Delta$Productivity $\sim$ & Ergonomics & 0.242 & 0.024 & 10.233 & 0.000  \\
 & DisasterPreparedness & 0.097 & 0.035 & 2.788 & 0.005 \\
 & adultCohabitants & 0.041 & 0.015 & 2.752 & 0.006 \\
 & disability & 0.124 & 0.049 & 2.513 & 0.012 \\
 & age & 0.070 & 0.032 & 2.220 & 0.026 \\
 & Fear & -0.002 & 0.009 & -0.204 & 0.838 \\
 & & & & & \\
$\Delta$Wellbeing $\sim$ & $\Delta$Performance & 0.822 & 0.045 & 18.361 & 0.000 \\
\noalign{\smallskip}\hline
\end{tabular}

\emph{Notes: converged after 97 iterations; Latent variables capitalized (e.g. Fear); direct measurements in camelCase (e.g. age, adultCohabitants)}
\end{table}

Inspecting the detailed SEM results (Table \ref{tab:SEM}) reveals numerous interesting patterns. Direct effects include: 
\begin{itemize}
    \item People who live with small children have significantly less ergonomic home offices. This is not surprising because the ergonomics scale included items related to noise and distractions. 
    \item Women tend to be more fearful. This is consistent with studies on the SARS epidemic, which found that women tended to perceive the risk as higher \citep{brug2004sars}. 
    \item People with disabilities are less prepared for disasters, have less ergonomic offices and are more afraid. 
    \item People who live with other adults are more prepared for disasters. 
    \item People who live alone have more ergonomic home offices.
    \item People who have COVID-19 or have family members, housemates or close friends with COVID-19 tend to be more afraid, more prepared, and have worse wellbeing since working from home. 
    \item People who are more isolated (i.e. not leaving home at all, or only for necessities) tend to be more afraid.   
\end{itemize}

Some indirect effects are also apparent, but are more difficult to interpret. For example, changes in productivity and wellbeing are closely related. Hypothesis H4 may be unsupported because change in productivity is mediating the effect of disaster preparedness on change in wellbeing. Similarly, Hypothesis H7 may not be unsupported because change in wellbeing is mediating the relationship between fear and change in productivity. Furthermore, control variables including gender, children and disability may have significant effects on wellbeing or productivity that are not obvious because they are mediated by another construct. Some variables have conflicting effects. For example, disability has not only a direct positive effect on productivity but also an indirect negative effect (through fear). So, is the pandemic harder on people with disabilities? More research is needed to explore these relationships.  

Above, we mentioned omitting language and country because SEM does not respond well to nominal categorical variables. We tried anyway, and both language and country were significant predictors for all latent variables, but, as expected, including so many binary dummy variables makes the model impossible to interpret. While our analysis suggests that country, language (and probably culture) have significant effects on disaster preparedness, ergonomics, fear, wellbeing and productivity, more research is need to understand the nature of these effects  (see Section \ref{sec:futureDirections}).

\subsection{Organizational support}
Table \ref{tab:support} shows participants' opinions of the helpfulness of numerous ways their organizations could support them. Several interesting patterns stand out from this data:
\begin{itemize}
    \item Only action \#1---paying developer's home internet charges---is perceived as helpful by more than half of participants and less than 10\% of companies appear to be doing that.
    \item The action most companies are taking (\#20, having regular meetings) is not perceived as helpful by most participants.
    \item There appears to be no correlation between things developers believe would help and things employers are actually doing.
    \item There is little consensus among developers about what their organizations should do to help them. 
\end{itemize}

\begin{table}
\caption{Organizational support actions in order of perceived helpfulness*}
\label{tab:support}

\begin{tabularx}{\columnwidth}{lXrr}
\hline\noalign{\smallskip}
\# & Action & Helpful & Following \\
\noalign{\smallskip}\hline\noalign{\smallskip}
1 & My organization will pay for some or all of my internet charges & 51.9\% & 9.8\%\\
2 & My organization will buy new equipment we need to work from home & 49.2\% & 30.9\%\\
3 & My organization is encouraging staff to use this time for professional training & 47.7\% & 24.3\%\\
4 & My organization has reassured me that they understand if my work performance suffers & 47.4\% & 40.5\%\\
5 & My organization is providing activities to occupy staff member's children & 46.4\% & 7.2\%\\
6 & My organization is sending food to staff working from home & 44.5\% & 4.0\%\\
7 & My organization is providing at-home exercise programs & 41.4\% & 15.8\%\\
8 & My organization has reassured me that I will keep my job & 40.2\% & 62.4\%\\
9 & My organization has reassured me that I can take time off if I'm sick or need to care for dependents & 40.1\% & 65.5\%\\
10 & My organization is improving documentation of its processes (e.g. how code changes are approved) & 37.4\% & 34.7\%\\
11 & My organization will pay for software we need to work from home & 36.8\% & 54.7\%\\
12 & My team is peer reviewing commits, change requests or pull requests (peer code review) & 36.5\% & 63.1\%\\
13 & I can (or could) take equipment (e.g. monitors) home from my workplace & 36.0\% & 73.9\%\\
14 & My organization has reassured me that I will continue to be paid & 34.7\% & 75.2\%\\
15 & My team uses a build system to automate compilation and testing & 34.3\% & 62.9\%\\
16 & Someone is keeping high priority work ready and our backlog organized & 33.1\% & 60.0\%\\
17 & My team has good work-from-home infrastructure (e.g. source control, VPN, remote desktop, file sharing) & 32.6\% & 86.4\%\\
18 & My team is having virtual social events (e.g. via video chat) & 32.1\% & 56.1\%\\
19 & My organization is encouraging staff to touch base regularly with each other & 30.8\% & 62.4\%\\
20 & My team is continuing to have regular meetings (e.g. via video chat) & 28.5\% & 88.9\%\\
21 & My team is avoiding synchronous communication (e.g. video chat) & 25.5\% & 14.3\%\\
22 & For most of the day, I work with an open video or audio call to some or all of my team & 23.3\% & 26.7\%\\
\noalign{\smallskip}\hline
\end{tabularx}

*number of respondents who indicated that this practice is or would be \emph{helpful} and number of respondents who indicated that their organizations are \emph{following} this recommendation (n=2225)
\end{table}

In hindsight, the structure of this question may have undermined discrimination between items. Future work could investigate a better selection of actions, and possibly ask participants for their ``top N'' items to improve reliability. Moreover, the helpfulness of actions may depend on where the participant lives; for example, in countries with a weaker social safety net, reassuring employees that they will keep their jobs, pay and benefits may be more important.  

\subsection{Summary interpretation}

This study shows that software professionals who are working from home during the pandemic are experiencing diminished emotional wellbeing and productivity, which are closely related. Furthermore, poor disaster preparedness, fear related to the pandemic, and poor home office ergonomics are exacerbating this reduction in wellbeing and productivity. Moreover, women, parents and people with disabilities may be disproportionately affected. In addition, dissensus regarding what organizations can do to help suggests that no single action is universally helpful; rather, different people need different kinds of support.

\section{Discussion}
\label{sec:discussion}

\subsection{Recommendations}
\label{sec:recommendations}
Organizations need to accept that expecting normal productivity under these circumstances is unrealistic. Pressuring employees to maintain normal productivity will likely make matters worse. Furthermore, companies should avoid making any decisions (e.g. layoffs, promotions, bonuses) based on productivity during the pandemic because any such decision may be prejudiced against protected groups. If a member of a protected group feels discriminated against due to low productivity at this time, we recommend contacting your local human rights commission or equivalent organization. 

Because productivity and wellbeing are so closely related, the best way to improve productivity is to help employees maintain their emotional wellbeing. However, no single action appears beneficial to everyone, so organizations should talk to their employees to determine what they need. 

Helping employees improve the ergonomics of their work spaces, in particular, should help. However, micromanaging foot positions, armrests and hip angles is not what we mean by ergonomics. Rather, companies should ask broad questions such as ``what do you need to limit distractions and be more comfortable?'' Shipping an employee a new office chair or noise cancelling headphones could help significantly.

Meanwhile, professionals should try to accept that their productivity may be lower and stop stressing about it. Similarly, professionals should try to remember that different people are experiencing the pandemic in very different ways---some people may be more productive than normal while others struggle to complete any work through no fault of their own. It is crucial to support each other and avoid inciting conflict over who is working harder.  

\subsection{Limitations and threats to validity}

The above recommendations should be considered in the context of the study's limitations. 

\paragraph{Sampling bias.} Random sampling of software developers is rare \citep{amir2018there} because there are no lists of all the software developers, projects, teams or organizations in the world or particular jurisdictions \citep{baltes2020sampling}. We therefore combined convenience and snowball sampling with a strategy of finding a co-author with local knowledge to translate, localize and advertise the questionnaire in a locally effective way. On one hand, the convenience/snowball strategy may bias the sample in unknown ways. On the other hand, our translation and localization strategy demonstrably increased sample diversity, leading to one of the largest and broadest samples of developers ever studied, possible due to a large, international and diverse research team. Any random sample of English-speaking developers is comparatively ethnocentric. The sample is not balanced, e.g. many more respondents live in Germany than all of southeast Asia, but we attempt to correct for that (see \textit{Internal Validity}, below).  

\paragraph{Response Bias.} Meanwhile, we found minimal evidence of response bias (in Section \ref{sec:validityAnalysis}). However, because the questionnaire is anonymous and Google Forms does not record incomplete responses, response bias can only be estimated in a limited way. Someone could have taken the survey more than once or entered fake data. 
Additionally, large responses from within a single country could skew the data but we correct for company size, language and numerous demographic variables to mitigate this.

\paragraph{Construct validity.} To enhance construct validity, we used validated scales for wellbeing, productivity, disaster preparedness and fear/resilience. Post-hoc construct validity analysis suggests that all four scales, as well as the ergonomics scale we created, are sound (Section \ref{sec:validityAnalysis}). However, perceived productivity is not the same as actual productivity. Although the HPQ scale correlates well with objective performance data in other fields \citep{kessler2003world}, it may not in software development or during pandemics. Similarly, we asked respondents their opinion of numerous potential mechanisms for organizational support. Just because companies are taking some action (e.g. having regular meetings) or respondents believe in the helpfulness of some action (e.g. paying their internet bills), does not mean that those actions will actually improve productivity or wellbeing.  

\paragraph{Measurement Validity.} There is much debate about whether 5- and 6-point responses should be treated as ordinal or interval. CFA and SEM are often used with these kinds of variables in social sciences despite assuming at least interval data. Some evidence suggests that CFA is robust against moderate deviations from normality, including arguably-ordinal questionnaire items \citep[cf.]{flora2004empirical}. We tend not to worry about treating data as interval as long as, in principle, the data is drawn from a continuous distribution. Additionally, due to a manual error, the Italian version was missing organizational support item 11: ``My team uses a build system to automate compilation and testing.'' The survey may therefore under-count the frequency and importance of this item by up to 10\%. 

\paragraph{Conclusion validity.} We use structural equation modeling to fit a theoretical model to the data. Indicators of model fit suggest that the model is sound. Moreover, SEM enhances conclusion validity by correcting for multiple comparisons, measurement error (by inferring latent variables based on observable variables), testing the entire model as a whole (instead of one hypothesis at a time) and controlling for extraneous variables (e.g. age, organization size). SEM is considered superior to alternative path modeling techniques such as partial least squares path modeling \citep{ronkko2013critical}. While a Bayesian approach might have higher conclusion validity \citep{furia2019bayesian}, none of the Bayesian SEM tools (e.g. \textsf{Blaavan}) we are aware of support ordered categorical variables. The main remaining threat to conclusion validity is overfitting the structural model. More research is needed to determine whether the model overstates any of the supported effects. 

\paragraph{Internal validity.} To infer causality, we must demonstrate correlation, precedence and the absence of third variable explanations. SEM demonstrates correlation. SEM does not demonstrate precedence; however, Wwe can be more confident in causality where precedence only makes sense in one direction. For example, having COVID-19 may reduce one's productivity, but feeling unproductive cannot give someone a specific virus. Similarly, it seems more likely that a more ergonomic office might make you more productive than that being more productive leads to a more ergnomic office. Meanwhile, we statistically controlled for numerous extraneous variables (e.g. age, gender, education level, organization size). However, other third-variable explanations cannot be discounted. Developers who work more overtime, for example, might have lower wellbeing, worse home office ergonomics, and reduced disaster preparedness. Other confounding variables might include individual differences (e.g. personality), team dynamics, organizational culture, family conflict, past medical history and wealth.

\subsection{Implications for researchers and future work}
\label{sec:futureDirections}
For researchers, this paper opens a new research area intersecting software engineering and crisis, disaster and emergency management. Although many studies explore remote work and distributed teams, we still need a better understanding of how stress, distraction and family commitments affect developers working from home \textit{during crises, bioevents and disasters}. More research is needed on how these events affect team dynamics, cohesion, performance, as well as software development processes and practices.

More specifically, the dataset we publish alongside this paper can be significantly extended. Abundant quantitative data is available regarding different countries, and how those countries reacted to the COVID-19 pandemic. Country data could be merged with our dataset to investigate how different contexts, cultures and political actions affect developers. For example, the quality of a country's social safety net may affect fear.  

Furthermore, the crisis continues. More longitudinal research is needed to understand its long-term effects on software professionals (e.g. burnout), projects (e.g. decreased velocity) and communities (e.g. trust issues). Research is also needed to understand how the crisis affects different kinds of professions. We focus on software developers because that is who software engineering research is responsible for, in the same way nursing researchers will study nurses and management researchers will study managers. Only comparing studies of different groups will reveal the extent to which our findings are specific to software professionals or generalizable to other knowledge workers. 

This study does not investigate typical software engineering practices (e.g. pair programming, mutation testing) or debates (e.g. agile methods vs. model-driven engineering) because we do not believe that a team's software development methodology is a key antecedent of pandemic-induced changes to productivity and wellbeing. Further research is needed to confirm or refute our intuition. 

\subsection{Lessons learned}

This study taught us two valuable lessons about research methodology. First, collaborating with a large, diverse, international research team and releasing a questionnaire in multiple languages \textit{with location-specific advertising} can generate a large, diverse, international sample of participants. 

Second, Google Forms should not be used to conduct scientific questionnaire surveys. It is blocked in some countries. It does not record partial responses or bounce rates, hindering analysis of response bias. URL parameter passing, which is typically used to determine how the respondent found out about the survey, is difficult. Exporting the data in different ways gives different variable orders, encouraging mistakes. Responses are recorded as (sometimes long) strings instead of numbers, overcomplicating data analysis. We should have used a research focused survey tool such as \textsf{LimeSurvey(.org)} or \textsf{Qualtrics(.com)}.

\section{Conclusion}
\label{sec:conclusion}

The COVID-19 pandemic has created unique conditions for many software developers. Stress, isolation, travel restrictions, business closures and the absence of educational, child care and fitness facilities are all taking a toll. Working from home under these conditions is fundamentally different from normal working from home. This paper reports the first large-scale study of how working from home during a pandemic affects software developers. It makes several key contributions:
\begin{itemize}
    \item evidence that productivity and wellbeing have declined;
    \item evidence that productivity and wellbeing are closely related; 
    \item a model that explains and predicts the effects of the pandemic on productivity and wellbeing;
    \item some indication that different people need different kinds of support from their organizations (there is no silver bullet here);
    \item some indication that the pandemic may disproportionately affect women, parents and people with disabilities.
\end{itemize}

Furthermore, this study is exceptional in several ways: (1) the questionnaire used previously validated scales, which we re-validated using both principal components analysis and confirmatory factor analysis; (2) the questionnaire attracted an unusually large sample of \validResponses responses; (3) the questionnaire ran in 12 languages, mitigating cultural biases; (4) the data was analyzed using highly sophisticated methods (i.e. structural equation modelling), which rarely have been utilized in software engineering research; (5) the study investigates an emerging phenomenon, providing timely advice for organizations and professionals; (6) the study incorporates research on emergency and disaster management, which is rarely considered in software engineering studies.  

We hope that this study inspires more research on how software development is affected by crises, pandemics, lockdowns and other adverse conditions. As the climate crisis unfolds, more research intersecting disaster management and software engineering will be needed. 

\section{Data Availability \label{sec:dataAvailability}}

A comprehensive replication package including our (anonymous) dataset, instruments and analysis scripts is stored in the Zonodo open data archive at \url{https://zenodo.org/record/3783511}.

\begin{acknowledgements}
This project was supported by the Natural Sciences and Engineering Research Council of Canada Grant RGPIN-2020-05001, the Government of Spain through project ``BugBirth'' (RTI2018-101963-B-100),  Dalhousie University and the University of Adelaide. Thanks to Brett Cannon, Alexander Serebrenik, Klaas Stol for their advice and support, as well as all of our pilot participants. Thanks also to all of the media outlets who provided complementary advertising, including DNU.nl, eksisozluk, InfoQ and Heise Online. Finally, thanks to everyone at \textit{Empirical Software Engineering} for fast-tracking COVID-related research. 
\end{acknowledgements}

%
%

\bibliographystyle{spbasic}      
\bibliography{bib}   

\end{document}